\documentclass{ws-procs975x65}

\newcommand{\chushi}[1]{}

\newcommand{\beq}{\begin{eqnarray}}
\newcommand{\eeq}{\end{eqnarray}}

\begin{document}

\title{One-Family Walking Technicolor in Light of LHC Run-II
}

\author{Shinya Matsuzaki}

\address{ Department of Physics, Nagoya University, Nagoya 464-8602, Japan.}
\address{ Institute for Advanced Research, Nagoya University, Nagoya 464-8602, Japan.}

\begin{abstract}
The LHC Higgs can be identified as the technidilaton, a composite scalar, 
arising as a pseudo Nambu-Goldstone boson for the spontaneous breaking
of scale symmetry in walking technicolor. 
One interesting candidate for the walking technicolor is the QCD with 
the large number of fermion flavors, involving the one-family model having 
the eight-fermion flavors. 
The smallness of the technidilaton mass can be ensured by 
the generic walking feature, Miransky scaling, and the presence of the ``anti-Veneziano limit" 
characteristic to the large-flavor walking scenario. 
To tell the standard-model Higgs from the technidilaton, one needs to wait for the precise estimate
of the Higgs couplings to the standard model particles, which is expected at the ongoing LHC-Run II. 
In this talk the technidilaton phenomenology in comparison with the LHC Run-I data
is summarized with the special emphasis placed on the presence of the anti-Veneziano limit supporting 
the lightness of technidilaton.  
Besides the technidilaton, the walking technicolor predicts
the rich particle spectrum such as technipions and technirho mesons,
arising as composite particles formed by technifermions.  
The LHC phenomenology of those technihadrons and the discovery channels 
are also discussed, which are smoking-guns of the walking technicolor,
to be accessible at the LHC-Run II. 

\end{abstract}

\bodymatter

\section{Introduction}

A Higgs boson with the mass about 125 GeV was discovered at the LHC Run I~\cite{Aad:2012tfa} 
and its coupling property has so far been almost consistent with the Higgs boson predicted in the standard model (SM). 
Yet the dynamical origin of the Higgs, related to the issue left in the SM (such as the naturalness problem)  
 has been uncovered, which is of great importance to be explored at the ongoing LHC Run II.

The dynamical origin of the electroweak symmetry breaking and the Higgs 
can elegantly be supplied by so-called technicolor (TC)~\cite{Weinberg:1975gm,Farhi:1980xs}.  
However, the technicolor based on the naive-scale up version of QCD (QCD-like technicolor) 
 has severely been disfavored by 
several inconsistencies with experiments. 
Most dramatically, it was ruled out by the recent discovery of the Higgs at LHC,
because of the absence of the light Higgs at around 125 GeV in the QCD-like technicolor.

In sharp contrast, 
the walking technicolor (WTC)~\cite{Yamawaki:1985zg,Bando:1986bg} 
predicts a light composite Higgs, which we call the technidilaton (TD).   
The TD  arises as a pseudo Nambu-Goldstone (NG) boson for the spontaneous breaking of 
the approximate scale symmetry of the WTC, triggered by technifermion condensation. 
Thus its lightness can potentially be ensured by the approximate scale symmetry inherent to the WTC.

One interesting candidate to realize the walking theory is one-family model with the number of techniflavors $N_F=8$~\cite{Dimopoulos:1979sp,Farhi:1980xs},  
classified into QCD with the large number of flavors.  
As it will turn out, actually, the large flavor nature of the walking 
provides us with a limit where the TD can be regarded as an exactly massless NG boson, 
analogously to the $\eta'$ in the Veneziano limit of QCD. 
That is what will be called ``anti-Veneziano limit".

The LHC signatures of the TD in the one-family model of WTC 
were studied~\cite{Matsuzaki:2011ie,Matsuzaki:2012gd,Matsuzaki:2012vc,Matsuzaki:2012mk,Matsuzaki:2012xx,Matsuzaki:2013fqa,Matsuzaki:2015sya}. 
It has been shown~\cite{Matsuzaki:2012vc,Matsuzaki:2013fqa,Matsuzaki:2015sya} that 
when the walking theory has the number of technicolor $N_C=4$,  
the coupling property of the 125 GeV TD is consistent with the LHC Higgs 
at the almost same level as the SM Higgs.

This talk summarizes the 125 GeV TD phenomenology at the LHC Run I and 
discusses future prospect in the Run II. 
The emphasis is also placed on several supports on the theoretical ground that the TD can indeed be as light as 
the 125 GeV boson due to the intrinsic feature of large flavor WTC. 
 Besides the TD, the walking technicolor predicts
the rich particle spectrum such as technipions and technirho mesons,
arising as composite particles formed by technifermions, 
The LHC phenomenology of those technihadrons and the discovery channels 
are also discussed, which are smoking-guns of the walking technicolor,
to be accessible at the LHC-Run II.

\section{Characteristic Features of Walking Technicolor}\label{WTC-TD}

\begin{figure}[t]
\begin{center}
\includegraphics[width=4.5cm]{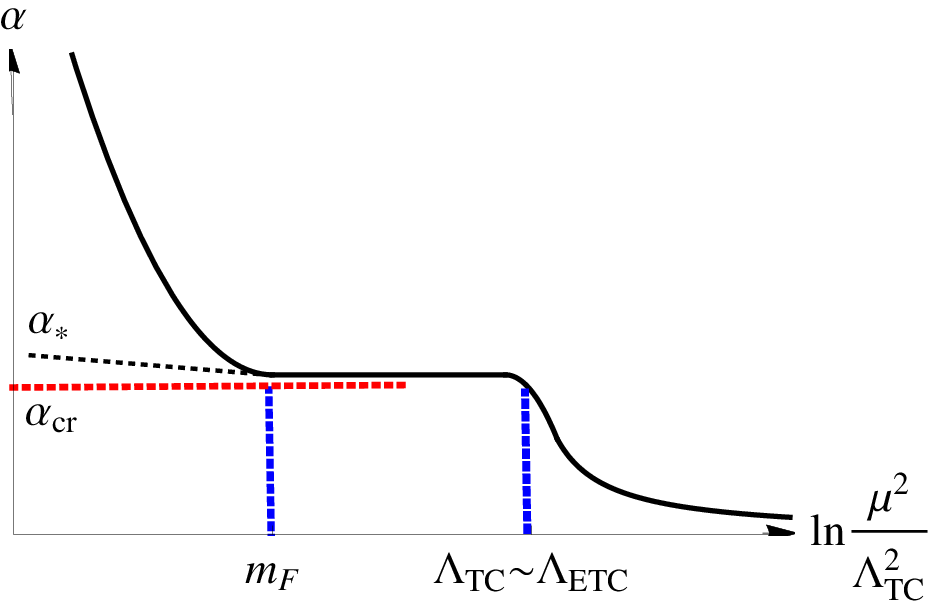} 
\hspace{15pt}
   \includegraphics[width=4.5cm]{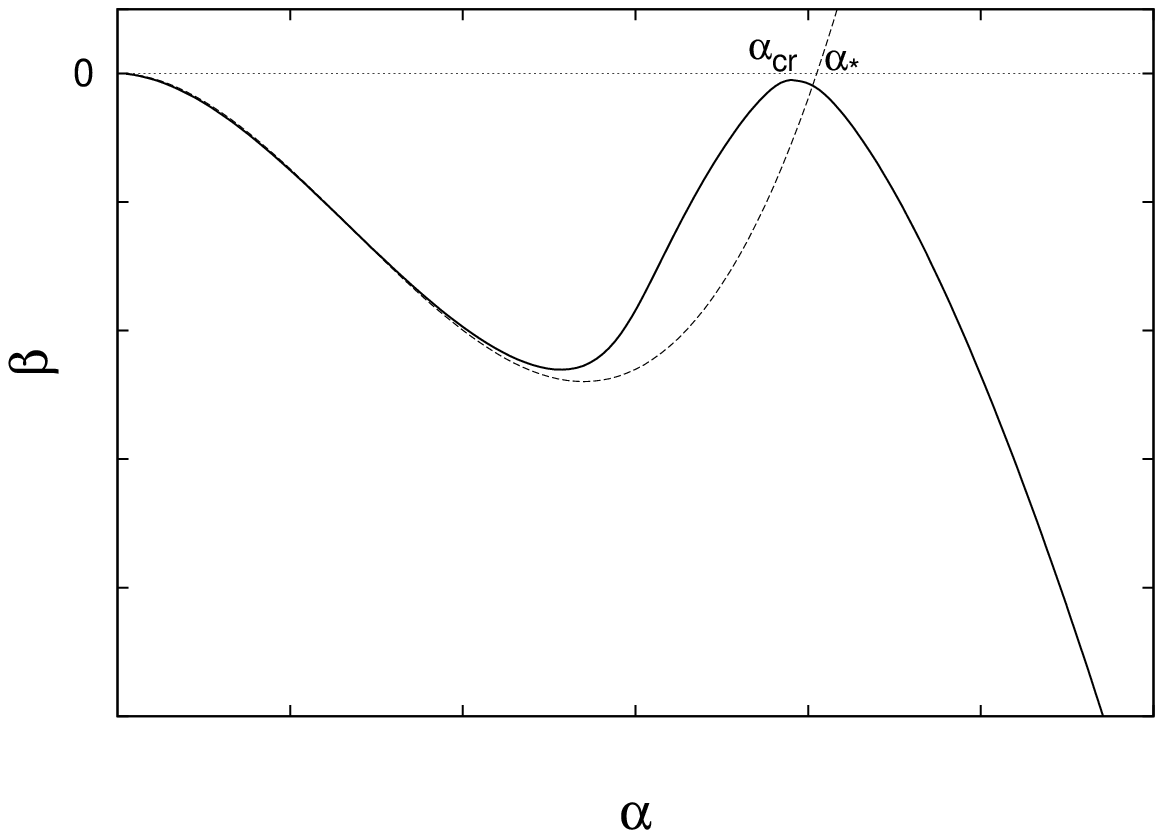} 
\vspace{15pt}
\caption{ A schematic picture of possible perturbative running coupling (left) and the beta function (right) in the region  $\alpha<\alpha_{\rm cr}$, in comparison with the nonperturbative region  $\alpha>\alpha_{\rm cr}$.
}
\label{beta:whole}
\end{center} 
 \end{figure}

In Fig.~\ref{beta:whole} a schematic view of the WTC is depicted  
in terms of the gauge coupling $\alpha$ (left panel) and its beta function $\beta(\alpha)$ (right panel). 
The walking, almost nonrunning region is literally defined as the domain where 
the running behavior of $\alpha$ looks like almost constant in scale (or the beta function $\beta$ looks almost zero in terms of the beta function $\beta$). 
The presence of walking region implies a pseudo infrared fixed point ($\alpha_*$), 
a la Caswell-Banks-Zack, based on the two-loop beta function in the large $N_f$ QCD~\cite{Caswell:1974gg}. 
During the walking region, the gauge coupling slowly reaches the critical coupling, 
which is slightly off from an infrared fixed point $\alpha_*$, 
$\alpha_{\rm cr}(< \alpha_*)$, where the chiral/electroweak symmetry is dynamically broken by technifermion condensation $\langle \bar{F}F \rangle \neq 0$ 
and hence technifermions get the dynamical mass $m_F$ on the order of TeV, 
where $F_\pi$ denotes the technipion decay constant associated with the chiral symmetry breaking.  
The walking regime ends at two edges:  
%in a wide range $(m_F< \mu < \Lambda_{\rm TC})$: 
below the infrared scale $m_F\sim {\cal O}({\rm TeV})$, technifermions decouple and hence the balance with technigluon 
contributions gets lost, leading to the one edge of walking.   
The other edge is above the ultraviolet scale $\Lambda_{\rm TC}(\sim 10^3-10^4$ TeV) the theory 
will be embedded into an extended technicolor (ETC)~\cite{Dimopoulos:1979es}.

The dynamical mass generation at $\alpha = \alpha_{\rm cr}$ 
is characterized by so-called Miransky scaling~\cite{Miransky:1984ef}, 
tied with the conformal phase transition~\cite{Miransky:1996pd}, 
\begin{equation} 
 m_F \sim \Lambda_{\rm TC} e^{-\frac{\pi}{\sqrt{\alpha/\alpha_{\rm cr} -1}}}
 \,, \qquad 
{\rm for} \qquad \alpha > \alpha_{\rm cr}
 \,. \label{Miransky}
\end{equation} 
This scaling property, i.e. the criticality actually supports the existence of the 
wide-range walking regime above $\alpha = \alpha_{\rm cr}$ in Fig.~\ref{beta:whole}, 
i.e., the large scale hierarcy $m_F \ll \Lambda_{\rm TC}$. 
Thus, the WTC realizes a typical technihadron mass scale $M_{\rm Had}$ 
on the order of electroweak/TeV scale, much smaller than $\Lambda_{\rm TC}$:  
\begin{equation} 
 M_{\rm Had} = {\cal O}({\rm EW}) \ll \Lambda_{\rm TC}
\,, 
\end{equation}
in sharp contrast to 
QCD where $M_{\rm Had} \sim \Lambda_{\rm QCD}$.

As seen from Fig.~\ref{beta:whole},  
the WTC possesses the (approximate) scale invariance ($\beta(\alpha) \simeq 0$ for $m_F < \mu < \Lambda_{\rm TC}$),   
which is spontaneously broken by the technifermion condensation/mass generation. 
This implies the presence of a (pseudo) Nambu-Goldstone boson (``dilaton") for the scale symmetry. 
which arises as a flavor-singlet composite scalar $\sim \bar{F}F$. 
This composite scalar is what we call the TD, technidilaton.

The mass of the TD is essentially provided by the ``nonperturbative" scale anomaly 
of the WTC: 
 according to the dynamical mass generation in Eq.(\ref{Miransky}), 
 the gauge coupling $\alpha$ is renormalized, starts running and hence  
the ``nonperturbative" beta function $\beta_{\rm NP}(\alpha)$ is generated~\cite{Bardeen:1985sm}: 
\begin{equation} 
\beta_{\rm NP}(\alpha) 
= \Lambda_{\rm TC} \frac{\partial \alpha}{\partial \Lambda_{\rm TC}}
= - \frac{2\alpha_{\rm cr}}{\pi} \left(\frac{\alpha}{\alpha_{\rm cr}}-1 \right)^{3/2}
\, \label{NP:beta}.  
\end{equation} 
 This induces the ``nonperturbative" scale anomaly: 
 \begin{equation}
  \partial_\mu D^\mu = \frac{\beta_{\rm NP}(\alpha)}{4 \alpha^2} \left( 
\alpha G_{\mu\nu}^2 \right)  \neq 0 
\,,  \label{NP:SA}
 \end{equation}
where $D_\mu$ denotes the dilatation current and $G_{\mu\nu}$ the field strength of technigluon field. 
Thus the TD becomes massive due to the nonperturbative scale anomaly~\cite{Yamawaki:1985zg,Bando:1986bg}. 
Note that this scale anomaly is induced by the fermion mass generation itself: 
the scale symmetry is spontaneously broken by the technifermion condensation, 
at the same time it is explicitly broken by the nonperturbative running, 
as in Eq.(\ref{NP:beta}), triggered by the technifermion mass generation itself.

The TD mass generation can be dictated by the partially conserved dilation current (PCDC) relation,  
\begin{equation} 
 \langle 0 | \theta_\mu^\mu  |0 \rangle 
= \frac{F_\phi^2 M_\phi^2}{4} 
\,, \label{PCDC}
\end{equation}
where $\theta_\mu^\mu= \partial_\mu D^\mu$ is the trace of the energy-momentum tensor as in Eq.(\ref{NP:SA}) and $F_\phi$ is the TD decay constant defined as 
\begin{equation} 
  \langle 0 | D_\mu(x)  |\phi(p) \rangle 
= - i p_\mu F_\phi e^{- i px}
\,. 
\end{equation} 
(The $F_\pi$ is not necessarily equal to $F_\pi$, can rather be larger, 
as it will turn out below. ) 
Since the scale anomaly is generated by the technifermion mass generation, 
one can find that the $\langle 0 | \theta_\mu^\mu  |0 \rangle$ generically scales like 
\begin{equation} 
 \langle 0 | \theta_\mu^\mu  |0 \rangle 
\sim N_C N_F m_F^4 
\,, 
\end{equation}
up to some loop factor dependent on details of nonperturbative computations. 
Combining this with the PCDC in Eq.(\ref{PCDC}), one arrives at 
a generic scaling law of the TD mass, 
\begin{equation} 
 M_\phi \sim \sqrt{N_C N_F} \frac{m_F^2}{F_\phi}
\,. \label{scaling:TDmass}
\end{equation}

There have been much progress of the WTC particularly on the light TD, 
not just in the ladder Schwinger-Dyson (SD) equation, but also in a  variety of approaches such as the ladder Bethe-Salpeter equation 
combined with the ladder SD equation~\cite{Harada:2003dc,Kurachi:2006ej}, 
the effective theory based on the scale-invariant chiral perturbation 
theory~\cite{Matsuzaki:2012vc,Matsuzaki:2013eva}, 
holographic method~\cite{Haba:2010hu,Matsuzaki:2012xx,Kurachi:2014xla}, and eventually, 
the first-principle calculation of the flavor-singlet scalar meson in the large $N_F$ QCD 
on the lattice \cite{Aoki:2014oha,Aoki:2013zsa,Fodor:2014pqa, Brower:2014ita}. 
In particular, it is remarkable that such a light flavor-singlet scalar meson as a candidate for the TD was observed in the lattice $N_F=8$ QCD \cite{Aoki:2014oha}, 
the theory shown to  have signatures of the lattice walking theory including 
the mass anomalous dimension $\gamma_m \simeq 1$ \cite{Aoki:2013xza, Appelquist:2014zsa,Hasenfratz:2014rna}. 
Note that $N_F=8$ (four weak-doublets)  corresponds to the ``one-family model'' \cite{Dimopoulos:1979sp,Farhi:1980xs}  which is the  most straightforward  model building 
 of  the ETC~\cite{Dimopoulos:1979es} as a standard way to give masses to 
the quarks and leptons. 
The one-family model of the WTC with $N_C=4$ is in fact best fit to the 125 GeV Higgs data~\cite{Matsuzaki:2012gd,Matsuzaki:2012vc,Matsuzaki:2012mk,Matsuzaki:2012xx}, which has been 
updated recently~\cite{Matsuzaki:2015sya} and will be summarized later,  
and is shown to be most natural for the ETC model building \cite{Kurachi:2015bva}.

Below we shall propose a novel understanding for the realization of the light TD: 
it is the presence of the ``anti-Veneziano limit'', intrinsic to the WTC based on 
large $N_F$ QCD. This observation has separately been described in the recent paper~\cite{Matsuzaki:2015sya}.

\section{Parametrically Light Technidilaton in Large $N_F$ Walking Technicolor}

 The key observation is that as long as the PCDC relation is satisfied,  
the TD as a pseudo NG boson has a vanishing mass in the anti-Veneziano limit, 
quite independently of the numerical details of nonperturbative calculations. 
 To see this one should note from the large $N_F$ and $N_C$ scalings that 
\begin{equation} 
 m_F \sim \frac{F_\pi}{\sqrt{N_C}} 
\,, \qquad 
 F_\phi \sim \sqrt{N_C N_F} m_F 
\,, 
\end{equation} 
and the $F_\pi$ is related to the electroweak scale $v_{\rm EW}\simeq 246$ GeV as 
$F_\pi = 1/\sqrt{N_F/2}\, v_{\rm EW}$. 
Then one immediately finds that the TD mass scaling in Eq.(\ref{scaling:TDmass}) implies 
\begin{equation} 
 \frac{M_\phi}{v_{\rm EW}} 
\sim \frac{1}{\sqrt{N_F N_C}}
\,. \label{TDmass:AV}
\end{equation} 
Now consider the anti-Veneziano limit where 
\begin{equation} 
 N_C \to \infty
\,, \qquad 
N_F \to \infty 
\,, \qquad 
{\rm with} 
\qquad 
r \equiv \frac{N_F}{N_C} \gg 1 
\,\qquad  
{\rm fixed}
\,. \label{a-V:limit}
\end{equation}
In this limit one readily gets $M_\phi/v_{\rm EW} \to 0$ in the chiral/electroweak broken phase with 
$v_{\rm EW}\simeq 246$ GeV fixed.   
Thus the TD as the pseudo NG boson  has a vanishing mass limit, though not exact massless point, 
  in the anti-Veneziano limit,  
  where the nonperturbative scale anomaly vanishes in units of $F_\phi$ via the PCDC in Eq.(\ref{PCDC}) 
  as a measure of the spontaneous symmetry breaking of the scale symmetry. 
  This is  similar to the $\eta^\prime$ meson in QCD, which is regarded as a pseudo NG boson whose mass, evaluated through the anomalous WT identity with the $U(1)_A$ anomaly,
   does vanish  in   the large $N_F$ and $N_C$ limit with $r=N_F/N_C$ fixed ($\ll 1$) (Veneziano limit): 
$M_{\eta^\prime}^2/F_\pi^2 \sim N_F/N_C^2 \rightarrow 0$, without the exact massless point. 
In this sense, the TD can be a vanishingly light pseudo NG boson at the same level as 
the $\eta'$ in QCD.

Actually, more suprising thing will happen when one employs the 
ladder approximation for the WTC with the nonrunning gauge coupling.  
(The details has been given in the full paper~\cite{Matsuzaki:2015sya}.)    
Then the fermion dynamical mass $m_F$ is related to the electroweak scale $v_{\rm EW}$ precisely through  
the Pagels-Stokar formula:  
\begin{equation} 
v_{\rm EW}^2=(246\,  {\rm GeV})^2 
\simeq  
\frac{N_F N_C}{4\pi^2} \, m_F^2  
\simeq  m_F^2 \left[\frac{N_F}{8}\frac{N_C}{4}\right]
\,,
\end{equation} 
%with $N_D (=N_F/2)$ being the number of the electroweak doublets. 
 From this and Eq.(\ref{TDmass:AV}) one can find 
a natural estimate of the TD mass for the one-family model with $N_F=8$ and $N_C=4$ 
to be 
\begin{equation} 
M_\phi = {\cal O} (m_F/2) ={\cal O} ( v_{\rm EW}/2) ={\cal O} (125 \, {\rm GeV})
\,, 
\end{equation}
in agreement with the LHC Higgs as the TD.

More precisely, one has 
\beq 
M_\phi^2 \simeq \left(\frac{v_{\rm EW}}{2}\right)^2 \cdot 
\left(  \frac{5\, v_{\rm EW}}{F_\phi} \right)^2 \cdot \left[\frac{8}{N_F}\frac{4}{N_C}\right]. 
\eeq  
%It was first pointed out in Ref. \cite{Matsuzaki:2012gd} that this ladder PCDC result accommodates the $125\, {\rm GeV}$ Higgs
%with $F_\phi = {\cal O}\, ( {\rm TeV})$ for the one-family model with $N_F=8$ and 
Remarkablly, it  
was shown~\cite{Matsuzaki:2012gd,Matsuzaki:2012vc,Matsuzaki:2012mk,Matsuzaki:2012xx} 
that the TD best fit to the current LHC data is realized when 
\beq
F_\phi \simeq 5\, v_{\rm EW} \simeq 1.25 \, {\rm TeV}\quad {\rm for}\, \quad M_\phi=125 \, {\rm GeV}\quad (N_F=8,\, N_C=4)
\label{Fphi}
\eeq
(See also the later discussions). With 
the fact that $v_{\rm EW}^2 \propto N_FN_C m_F^2 \sim F_\phi^2$, the result reflects
the generic scaling: 
\beq
\frac{M_\phi}{v_{\rm EW}} \sim \frac{M_\phi}{F_\phi} \sim \frac{m_F}{F_\phi} \sim \frac{1}{\sqrt{N_F N_C}} \rightarrow 0, 
\eeq
in the anti-Veneziano limit.

\section{The low-energy effective model for the TD}

The effective theory for the TD involving the SM particles are constructed based on 
the scale-invariant chiral perturbation theory~\cite{Matsuzaki:2012vc,Matsuzaki:2013eva} with the scale anomaly terms  
supplied properly to the underlying WTC.

The chiral/electroweak and scale invariant Lagrangian thus takes the form~\cite{Matsuzaki:2012vc}
\begin{equation}
{\cal L}_{\rm inv} = \frac{v_{\rm EW}^2}{4} \chi^2 {\rm tr}[D_\mu U^\dag D^\mu U] + {\cal L}_{\rm kin}(\chi) 
\,, \label{inv:L}
\end{equation}
where $\chi=e^{\phi/F_\phi}$ is a nonlinear base of the scale symmetry, which parametrizes the TD field $\phi$ with the decay constant $F_\phi$ 
and has the scale dimension 1; $D_\mu U= \partial_\mu U - i W_\mu U + i U B_\mu$ with the $SU(2)_W$ and $U(1)_Y$ gauge fields 
$W$ and $B$; ${\cal L}_{\rm kin}(\chi)$ denotes the scale invariant kinetic term of TD and $U$ the usual chiral field parameterizing 
the (eaten) Nambu-Goldstone boson fields $\pi$ as $U=e^{2 i \pi/v_{\rm EW}}$; 
The $|D_\mu U|^2$  term gives the TD couplings to massive weak bosons: 
\begin{equation} 
g_{\phi WW/ZZ} = \frac{2 m_{W/Z}^2}{F_\phi} 
 \,. \label{td:WW}
\end{equation}

As seen in Eqs.(\ref{NP:beta}) and (\ref{NP:SA}), 
the scale symmetry is actually broken explicitly as well as spontaneously  by 
dynamical  mass generation of technifermions, which   
 has to be respected also in the nonlinear realization~\cite{Matsuzaki:2012vc}. 
 Such explicit breaking effects arise in the TD Yukawa couplings to the SM fermions, 
which reflect underlying ETC-induced four-fermion terms, and couplings to QCD gluons and photons 
related to the scale anomaly in the SM gauge sector. 
In order to incorporate these effects into the scale-invariant Lagrangian, 
we introduce a spurion field $S$ having 
the scale dimension $1$ coupled to the SM fermions, digluon $gg$ and diphoton $\gamma\gamma$ 
in such a way that~\cite{Matsuzaki:2012vc} 
\begin{eqnarray} 
{\cal L}_{S} 
&=& 
- m_f \left( \left( \frac{\chi}{S} \right)^{2-\gamma_m} \cdot \chi \right) \bar{f} f 
\nonumber \\ 
&& 
+  \log \left( \frac{\chi}{S} \right) \left\{ 
\frac{\beta_F(g_s)}{2g_s} G_{\mu\nu}^2 
+ 
\frac{\beta_F(e)}{2e} F_{\mu\nu}^2 
\right\}  
\,, \label{Lag:S}
\end{eqnarray}
where $G_{\mu\nu}$ and $F_{\mu\nu}$ respectively denote the field strengths for QCD gluon and photon fields; 
$g_s$ and $e$ are the QCD gauge and electromagnetic couplings, respectively; 
$\beta_F$s are  the beta functions only including the technifermion 
loop contributions.

The TD Yukawa coupling to the SM $f$-fermion arises from the first line of Eq.(\ref{Lag:S}) as~\cite{Bando:1986bg} 
\begin{equation} 
 g_{\phi ff} =  \frac{(3-\gamma_m) m_f}{F_\phi} 
\,, \label{td:yukawa}
\end{equation}
along with scale dimension of technifermion bilinear operator 
$(3-\gamma_m)$. 
The anomalous dimension $\gamma_m \simeq 1$ in WTC, which is crucial to
obtain the realistic mass of the SM fermions of the first and the second generations without suffering from the 
flavor-changing neutral current problems.  
However it was known for long time that it is not enough for the mass of the third-generation 
SM $f$-fermions like $t, b, \tau$: 
A simplest resolution would be the strong ETC model~\cite{Miransky:1988gk} 
having much larger anomalous dimension $1<\gamma_m <2$ due to the strong effective four-fermion coupling
from the ETC dynamics 
in addition to the walking gauge coupling.  
Here we take $\gamma_m \simeq$ 2, i.e., $(3-\gamma_m) \simeq 1$, 
as in the strong ETC model 
for the third-generation SM $f$-fermions like $t, b, \tau$  
which are relevant to the current LHC data.

In addition, 
the TD potential should be included so as to 
reproduce the PCDC relation Eq.(\ref{PCDC}) in the underlying walking theory: 
\begin{equation} 
V(\chi)= \frac{F_\phi^2}{4} m_\phi^2 \chi^4 \left( \log \frac{\chi}{S} - \frac{1}{4}\right)
\,. \label{TD:potential}
\end{equation}

Here we stress remark on  stability of the light TD mass against radiative corrections.  
As a pseudo Nambu-Goldstone boson of scale invariance, the quadratic divergence is suppressed by the 
scale invariance for the walking regime $m_F(\simeq v_{\rm EW}) < \mu <\Lambda_{\rm TC}(\sim \Lambda_{\rm ETC})$. 
The scale symmetry breaking in the ultraviolet region $\mu>\Lambda_{\rm TC}$ has no problem thanks to 
the naturalness as usual just as in QCD where the  theory has  only logarithmic divergences. 
Only possible source of the scale symmetry violation is from an effective theory for $\mu < m_F(\simeq v_{\rm EW})$.   
See Fig.~\ref{beta:whole}.

Now it turns out that the TD mass is stable against the feedback effects of the ETC through particularly the top quark loop,
because of the large $F_\phi \simeq 5 v_{\rm EW}$: 
below $\mu=m_F(\simeq v_{\rm EW})$ 
the dominant corrections to the TD mass $M_\phi$ come from the SM top quark and TD self-loops.  
These can be estimated from the effective Lagrangian in Eqs.(\ref{inv:L}) and (\ref{Lag:S}) 
including the SM sector and ETC effects, which are estimated to
be~\cite{Matsuzaki:2012vc,Matsuzaki:2015sya}
\beq
\frac{\delta M_\phi^2|_{\phi^4}}{M_\phi^2} 
& \simeq & 
 24 \frac{m_F^2}{(4\pi F_\phi)^2} \simeq 6\times 10^{-3}
\,, \nonumber \\ 
\frac{\delta M_\phi^2|_{\rm ETC/Yukawa}}{M_\phi^2} 
& \simeq & 
12 (3-\gamma_m)^2 \frac{m_F^2}{(4\pi F_\phi)^2} \frac{m_t^2}{M_\phi^2} 
 \nonumber \\  
& \simeq& (3-\gamma_m)^2 \frac{\delta M_\phi^2|_{\phi^4}}{M_\phi^2} 
\,, \eeq
where the cutoff has been set to $m_F\simeq v_{\rm EW}$.  
These yield  
\begin{equation}
\delta M_\phi^2/M_\phi^2 =0.01 \, (\gamma_m=2) \, - 0.03 \, (\gamma_m=1) 
\, . 
\end{equation} 
Thus the 125 GeV TD mass is fully stable against the radiative corrections, in contrast to 
the unnatural SM Higgs case.

In passing, the TD potential in Eq.(\ref{TD:potential}) 
written in terms $\chi=e^{\phi/F_\phi}$ is rewritten in the TD field $\phi$ as~\cite{Matsuzaki:2012vc}
\begin{equation} 
V(\phi) =-{\cal L}^{S}_{(2){\rm anomaly}} 
= - \frac{M_\phi^2 F_\phi^2}{16} +\frac{1}{2}M_\phi^2\,\phi^2 +\frac{4}{3} \frac{M_\phi^2}{F_\phi} \,\phi^3 
+ 2 \frac{M_\phi^2}{F_\phi^2}\, \phi^4 
+ \cdots 
\,. 
\label{dilatonpotential}
\end{equation}
It is remarkable to notice that in the anti-Veneziano limit Eq.(\ref{a-V:limit})  
the TD self couplings (trilinear and quartic couplings) are highly suppressed:
\beq
\frac{4}{3} \frac{M_\phi^2}{F_\phi} \sim \frac{1}{\sqrt{N_F N_C}}\,,\quad 
2 \frac{M_\phi^2}{F_\phi^2} \sim \frac{1}{N_F N_C}
\eeq
by $M_\phi/F_\phi \sim 1/\sqrt{N_F N_C}$ and $M_\phi \sim N_F^0 N_C^0$. 
It is also interesting to numerically compare the TD self couplings for 
the one-family model ($N_F=8, N_C=4)$ having $v_{\rm EW}/F_\phi \simeq 1/5$ 
with 
the  self couplings of the SM Higgs with $m_h=M_\phi$, by making the ratios: 
\begin{eqnarray} 
\frac{g_{\phi^3}}{g_{h_{\rm SM}^3}}\Bigg|_{M_\phi=m_h} 
&=& \frac{\frac{4 M_\phi^2}{3 F_\phi}}{\frac{m_h^2}{2 v_{\rm EW}}} \Bigg|_{M_\phi=m_h}
\simeq \frac{8}{3} \left( \frac{v_{\rm EW}}{F_\phi}\right) \simeq 0.5 
\,, \nonumber \\ 
\frac{g_{\phi^4}}{g_{h_{\rm SM}^4}} \Bigg|_{M_\phi = m_h} 
&=& \frac{\frac{2 M_\phi^2}{F_\phi^2}}{\frac{m_h^2}{8 v_{\rm EW}^2}}\Bigg|_{M_\phi=m_h} 
= 16 \left( \frac{v_{\rm EW}}{F_\phi} \right)^2 \simeq 0.6 
\,. 
\label{selfcouplings:0}
\end{eqnarray}
This shows that the TD self couplings, although generated by the strongly coupled interactions, are even smaller than those of the 
SM Higgs, a salient feature of the approximate scale symmetry in the ant-Veneziano limit. 
This is in sharp contrast to 
the widely-believed folklore, ``Strong coupling solutions like Technicolor tend to lead to a strongly coupled Higgs''~\cite{Seiberg}.

\section{LHC Higgs vs. Technidilaton in One-Family Model with $N_C=4$ and $N_F=8$}

  One finds from the effective Lagrangian 
that the TD couplings to the SM gauge bosons and the SM fermions can just be obtained by 
scaling from the SM Higgs as $v_{\rm EW} \to F_\phi$~\cite{Matsuzaki:2012vc,Matsuzaki:2012mk}:  
\begin{eqnarray} 
  \frac{g_{\phi WW/ZZ}}{g_{ h_{\rm SM} WW/ZZ }}   
 &=&  \frac{g_{\phi ff}}{g_{h_{\rm SM} ff}}  \,
\quad ({\rm for} \quad f=t,b,\tau) 
\nonumber \\ 
  &=& \frac{v_{\rm EW}}{F_\phi} \,\quad \left[ \simeq \frac{1}{5} \ll 1\quad \left(N_F=8\,, N_C=4 \right)\right]
  \,.  \label{scaling}
\end{eqnarray} 
On the other hand, in the one-family model with $N_F=8$  
the couplings to digluon and diphoton 
include the colored/charged 
techni-fermion loop contributions along with a factor $N_C$~\cite{Matsuzaki:2012vc,Matsuzaki:2012mk},  
\begin{eqnarray} 
 {\cal L}_{\rm eff}^{\gamma\gamma,g g} =\frac{\phi}{F_\phi} \left\{ 
 \frac{\beta_F(g_s)}{2g_s} G_{\mu\nu}^2 + \frac{\beta_F(e)}{e} F_{\mu\nu}^2 
\right\}\,,\\
 \beta_F(g_s) = \frac{g_s^3}{(4\pi)^2} \frac{4}{3} N_C \,,\quad
\nonumber  
  \beta_F(e) = \frac{e^3}{(4\pi)^2} \frac{16}{9} N_C 
\,,   \label{betas}
\end{eqnarray} 
where the beta functions have been evaluated at the one-loop level. 
Thus one finds  
the scaling from the SM Higgs~\cite{Matsuzaki:2012mk},  
%(Detailed formulae are given in the Appendix of Ref.\cite{Matsuzaki:2012vc}),  
 \begin{eqnarray} 
\frac{g_{\phi gg}}{g_{h_{\rm SM} gg}} 
&\simeq & 
\frac{v_{\rm EW}}{F_\phi} 
\cdot 
\left( 1 + 2 N_C \right)   \,,
\nonumber \\ 
\frac{g_{\phi \gamma\gamma}}{g_{h_{\rm SM} \gamma\gamma}} 
&\simeq & 
\frac{v_{\rm EW}}{F_\phi} 
\cdot 
 \left( \frac{63 -  16}{47} - \frac{32}{47} N_C \right)  
\,,  \label{g-dip-dig}
\end{eqnarray} 
where in estimating the SM contributions  
we have incorporated only the top (the terms of 1 and 16/47 for $gg$ and $\gamma\gamma$ rates, respectively) and the 
$W$ boson (the term of 63/47 for $\gamma\gamma$ rate) loop contributions.  

The scaling laws for the TD couplings to 
the SM fermions and weak bosons in Eq.(\ref{scaling}) 
are similar to those for the couplings of other dilatons, or radions~\cite{Goldberger:2007zk}, 
while the TD couplings to diphoton and digluon in Eq.(\ref{g-dip-dig}) 
are significantly differrent in a sense that  these couplings 
include contributions beyond the SM (technifermion loop contributions), 
in contrast to other types of dilatons/radions.

 In Table~\ref{tab:BR} the branching fractions for relevant decay channels of the TD at 125 GeV 
are listed in the case of 
 the one-family model with $N_C=4$.  Note that the total width $\Gamma_{\rm tot}=1.15\, {\rm MeV}$  is smaller than the SM Higgs, which 
 reflects the weaker couplings than the those of the latter, in contrast to the widely spread folklore. 
 
\begin{table} 
\tbl{ The TD branching ratios at 125 GeV in the one-family model with $N_C=4$. The total width is also given.}
{
\begin{tabular}{|c|c|c|c|c|c|c|c|c|}
\hline 
%\hspace{15pt}
BR[\%] 
%\hspace{15pt}
 &
%\hspace{15pt}
 $gg$ 
%\hspace{15pt}
& 
%\hspace{15pt}
 $bb$ 
%\hspace{15pt}
 & 
%\hspace{15pt}
 $WW$ 
%\hspace{15pt}
& 
%\hspace{15pt}
$ZZ$ 
%\hspace{15pt}
 & 
%\hspace{15pt}
 $\tau\tau$ 
%\hspace{15pt}
 & 
%\hspace{15pt}
 $\gamma\gamma$ 
%\hspace{15pt}
&
%\hspace{15pt}
 $Z \gamma$ 
%\hspace{15pt}
& 
%\hspace{15pt}
$\mu\mu$ 
%\hspace{15pt}
 \\  
\hline \hline  
$\Gamma_{\rm tot}=1.15\, {\rm MeV}$ 
& 75.1 & 19.6 & 3.56 & 0.38 & 1.19 & 0.068 & 0.0048 & 0.0042 \\ 
\hline 
\end{tabular}
} 
\label{tab:BR}
\end{table}

 Calculating the signal strengths for the LHC production categories (gluon gluon fusion (ggF), vector boson fusion (VBF), 
vector boson associate production (VH) and top associate production (ttH)), 
\begin{equation} 
 \mu^{i}_{X_1X_2} = \frac{\sigma^i_{\phi} \times {\rm BR}(\phi \to X_1X_2)}{\sigma^i_{\rm h_{\rm SM}} \times {\rm BR}(h \to X_1X_2)}
 \,, 
\end{equation} 
as a function of the overall coupling $v_{\rm EW}/F_\phi$ for given the number of $N_C$, 
we may fit the $\mu^i_{X_1X_2}$ to the latest data on the Higgs coupling measurements~\cite{ATLAS:2015:Higgs%ATLAS:2013oma,Aad:2013wqa,ATLAS:2013wla,ATLAS-tautau,TheATLAScollaboration:2013lia,CMS:ril,CMS:xwa,Chatrchyan:2013iaa,CMS:2013yea,Chatrchyan:2014nva,Chatrchyan:2013zna
}.  
to determine the best-fit value of $v_{\rm EW}/F_\phi$. 
The result of the goodness of fit is shown in Table~\ref{TDfit}, which updates the previous analysis~\cite{Matsuzaki:2012mk}.  
 The Table~\ref{TDfit} shows that the TD in the one-family model with $N_C=4$ is favored by the current LHC Higgs data 
 as much the same level as the SM Higgs. 
Remarkably, the best fit value $[v_{\rm EW}/F_\phi]_{\rm best}\simeq 0.2$, i.e. $F_\phi \simeq 5 v_{\rm EW}$ for $N_C=4$ 
is in excellent agreement with 
the ladder estimate of the TD mass $\simeq 125$ GeV in Eq.(\ref{Fphi})!

In Table~\ref{mu-vals} 
we also make a list of the predicted signal strengths for each production category 
for the best fit value of $v_{\rm EW}/F_\phi \simeq 0.23$ in the case with $N_C=4$, 
along with the latest result reported from the ATLAS and CMS experiments~\cite{ATLAS:2015:Higgs}. 
Note the TD signal strengths in the dijet category (VBF), which involves the contamination by about 30\% from the ggF + gluon jets, 
$gg \to \phi + gg$. The contribution from the ggF is highly enhanced compared to the SM Higgs case, 
due to the extra techni-quark loop contribution, which compensates the overall suppression by the direct 
VBF coupling $v_{\rm EW}/F_\phi \simeq 0.2$ to lift the event rate up to be comparable to the SM Higgs case. 
%(The detailed estimate of the ggF contamination is given in Appendix~\ref{dijet}.) 
Note also the suppression of the VH-$b\bar{b}$-channel, which would be the characteristic signature of the TD 
to be distinguishable from the SM Higgs. More data from the upcoming LHC Run-II will draw a conclusive answer 
to whether or not the LHC Higgs is the SM Higgs, or the TD.

\begin{table} 
\tbl{The best fit values of $v_{\rm EW}/F_\phi$ for the one-family model with 
$N_C=3,4,5$ displayed together with 
the minimum of the $\chi^2$ ($\chi^2_{\rm min}$) normalized by the degree of freedom. 
Also has been shown in the last column the case of the SM Higgs corresponding to $N_C=0$ and $v_{\rm EW}/F_\phi=1$. }
{
\begin{tabular}{|c|c|c|}
\hline 
\hspace{10pt} $N_C$ \hspace{10pt} &  
\hspace{10pt} $[v_{\rm EW}/F_\phi]_{\rm best}$ \hspace{10pt} & 
\hspace{10pt}$\chi^2_{\rm min}/{\rm d.o.f.}$ \hspace{10pt} \\ 
\hline\hline 
3 & 0.27 & $ 25/17\simeq 1.5$  \\ 
\hline 
4 & 0.23 & $ 16/17 \simeq 0.92$\\ 
\hline 
5 & 0.17 & $ 32/17 \simeq 2.0 $ \\ 
\hline  
\hline 
0\,[\rm SM Higgs] & 1 & 8.0/18 $\simeq 0.44$ \\ 
\hline 
\end{tabular}
} 
\label{TDfit}
\end{table}

\begin{table}[ht] 
\tbl{
The predicted signal strengths of the TD with $v_{\rm EW}/F_\phi=0.23$ in the case of the one-family model with 
$N_C=4$. The numbers in the parentheses  correspond to the amount estimated without contamination from 
the ggF process. 
Also have been displayed the latest data on the Higgs coupling measurements reported from the 
ATLAS and CMS experiments~\cite{ATLAS:2015:Higgs}.  
}
{
\begin{tabular}{|c|c|c|} 
 \hline 
\hspace{10pt} 
TD signal strengths $(v_{\rm EW}/F_\phi=0.23, N_C=4)$ 
\hspace{10pt} 
& 
\hspace{10pt} 
ATLAS 
\hspace{10pt} 
& 
\hspace{10pt} 
 CMS  
 \hspace{10pt} 
\\  
 \hline \hline 
$\mu_{\gamma\gamma}^{\rm ggF} \simeq 1.4$ & $1.32 \pm 0.38$ & $1.13 \pm 0.35$ \\ 
\hline 
$\mu_{ZZ}^{\rm ggF} \simeq 1.0$ & $1.7 \pm 0.5$ & $0.83 \pm 0.28$\\ 
\hline a
$\mu_{WW}^{\rm ggF} \simeq 1.0$ & $0.98 \pm 0.28$ & $0.72 \pm 0.37$ \\ 
\hline 
$\mu_{\tau\tau}^{\rm ggF} \simeq 1.0$ & $2.0 \pm 1.4$ & $1.1 \pm 0.46$ \\ 
\hline 
\hline 
$\mu_{\gamma\gamma}^{\rm VBF} \simeq 0.87$ (0.019) & $0.8 \pm 0.7$ & $1.16 \pm 0.59$ \\ 
\hline 
$\mu_{ZZ}^{\rm VBF} \simeq 0.61$ (0.014) & $0.3 \pm 1.3$  & $1.45 \pm 0.76$ \\ 
\hline 
$\mu_{WW}^{\rm VBF} \simeq 0.61$ (0.014) &  $1.28 \pm 0.51$ & $0.62 \pm 0.53$ \\ 
\hline 
$\mu_{\tau\tau}^{\rm VBF} \simeq 0.61$ (0.014) &  $1.24 \pm 0.57$ & $0.94 \pm 0.41$ \\ 
\hline 
$\mu_{bb}^{\rm VH} \simeq 0.014$ &  $0.52 \pm 0.40$ & $1.0 \pm 0.50$ \\ 
\hline 
\end{tabular}
} 
\label{mu-vals}
\end{table}

The ATLAS and CMS have made a plot of the LHC Higgs couplings to the SM particles against the SM particle masses~\cite{SM-Higgs-Yukawa}, 
shown that the LHC Higgs couplings to fermions have aligned very well with the SM Higgs boson
properties. The plot has been made by assuming no contributions beyond the SM in loops, 
i.e., no contributions beyond SM to diphoton and digluon couplings. 
However, as explicitly seen from Eq.(\ref{betas}), the technidilaton couplings to 
diphoton and digluon significantly include the terms beyond the SM, technifermion contributions 
charged under the $U(1)_{\rm em}$ or QCD color. 
In this respect, such a plot cannot be applied to the technidilaton. 
In fact, the successful consistency with the LHC Higgs coupling measurement, as shown in Table~\ref{mu-vals}, 
is due to those beyond SM contributions, which especially enhance the ggF production cross section, 
balanced by the overall suppression due to the coupling $F_\phi$ larger than $v_{\rm EW}$ 
by a factor of 5.

\section{Other Walking Technihadrons: Technipions and Technirhos}

In addition to the TD, the WTC predicts 
the rich spectrum. 
Such typical technihadrons involve the technipions, technirhos (and techni$a_1$s.)  
In the case of the one-family model having the chiral $SU(8)_L \times SU(8)_R$ symmetry 
broken down to the vectorial $SU(8)_V$, the technipion spectra can be classified by 
the SM gauge charges, as well as the technirhos and techni-$a_1$s.

\subsection{Walking Technipions}

Technipion masses are all from explicit breaking outside of the WTC sector, 
i.e, SM gauge interactions and ETC gauge interactions. 
One can estimate the technipion masses in the WTC, 
based on the first order perturbation of the explicit chiral symmetry breaking by the ``weak gauge couplings'' of 
SM gauge interactions and the ETC gauge interactions (Dashen's formula).
This is the same strategy as the QCD estimate of the $\pi^+ - \pi^0$ mass difference.  
 It turns out~\cite{Harada:2005ru,Jia:2012kd,Kurachi:2014xla} that 
the technipion masses are enhanced through the chiral condensate by the anomalous dimension as 
$(Z_m^{-1})^2 \sim (\Lambda/m_F)^{2 \gamma_m}$~\cite{Holdom:1981rm,Yamawaki:1982tg,Yamawaki:1985zg}, 
to be on the order of TeV scale: 
\begin{equation} 
 M_{\pi} \sim {\cal O}({\rm TeV})
\,. \label{TPmass}
\end{equation} 
It is striking that 
although the explicit chiral symmetry breakings are formally
very small due to the ``weak gauge couplings'',  
the nonperturbative contributions from the WTC sector lift all the technipions masses to the TeV region so that they all lose the nature of the ``pseudo NG bosons''.    
This is actually a universal feature of the dynamics with large anomalous dimension, ``amplification of the symmetry violation''\cite{Yamawaki:1996vr}.  
This amplification effect should not be confused with that of  the pseudo NG boson mass due to the technifermion bare mass effects, 
like the pion mass due to the current quark mass,
$F_\pi^2 m_\pi^2 = 2 m \langle \bar \psi \psi \rangle$, 
which are not amplified by the large anomalous dimension, 
since the mass operator times the mass is totally invariant against the renormalization. 
(In the actual technicolor model, all the technifermions are set to be
 exactly massless and such a type of explicit breaking is not considered anyway.)

Note that although the left-over light spectra are just three exact NG bosons absorbed into $W/Z$ bosons, 
our theory with $N_F \gg 2$ in the anti-Veneziano limit is completely  different from the model with 
massless flavors $N_f=2$ where the symmetry  breaking is  $SU(2)_L \times SU(2)_R/SU(2)_V$. 
In fact, even though all the NG bosons, other than the three exact NG bosons to be absorbed into $W,Z$ bosons, 
are massive and decoupled from the low energy physics, 
they are composite of the linear combinations of all the $N_F$ technifermions, not just 2 of them.

Including the walking technipions one can discuss the LHC phenomenology 
based on the scale-invariant chiral perturbation theory described by the Lagrangian in Eq.(\ref{inv:L})~\cite{Jia:2012kd}.   
The LHC-Run I limits on the 60 massive technipions were evaluated to be~\cite{Jia:2012kd,Kurachi:2014xla} 
\begin{eqnarray}  
\begin{array}{llll} 
  \textrm{color-octet technipions ($\theta$)}: 
& M_\theta >  1.5 - 1.6 \,{\rm TeV} 
& \hspace{5pt} \textrm{from} \hspace{5pt}
& \theta \to t\bar{t}  
\\ 
  \textrm{color-triplet technipions ($T$)}: 
& M_T >  1.0 - 1.1 \,{\rm TeV} 
& \hspace{5pt} \textrm{from} \hspace{5pt}
& \textrm{leptoquark search}  
\\  
  \textrm{color-singlet technipions ($P$)}: 
& M_P > 850 \,{\rm GeV} 
& \hspace{5pt} \textrm{from} \hspace{5pt}
& \theta \to t\bar{t}   
\end{array} 
\,. \nonumber 
\end{eqnarray}  
More data from the Run II will thus 
reveal if walking technipions with the mass as large as ${\cal O}({\rm TeV})$ are present.

\subsection{Non-pseudo NG-technihadrons: walking technirhos} 

In contrast to the TD and technipions,  
all the non-NG boson technihadrons, such as the technirho, techni-$a_1$, etc., have 
no constraints from the PCDC as the explicit breaking of the scale symmetry but do have 
constraints from the spontaneous breaking of the scale symmetry, 
so that they should have masses on the scale of  spontaneous breaking  of the scale symmetry, 
characterized by $F_\phi$ much larger than $2 m_F$ of the naive nonrelativistic quark model picture:
   \beq
 M_\rho, M_{a_1}, \cdots = {\cal O} ({\rm TeV's}) > {\cal O} (F_\phi) \gg 2 m_F \gg M_\phi.
 \eeq
 In fact, the infrared conformal physics of the WTC should  be 
described by the low-lying composite fields as effective fields, 
in a way to realize all the symmetry structure of the underlying theory. 
 
  Such an effective theory of WTC is constructed as a straightforward extension of scale-invariant chiral perturbation 
theory~\cite{Matsuzaki:2012vc,Matsuzaki:2013eva}, i.e,, 
the scale-invariant version \cite{Kurachi:2014qma} of the Hidden Local Symmetry (HLS) model \cite{Bando:1984ej,Harada:2003jx},  
(the ``s-HLS model''),
 where  the technirho mass terms have the {\it scale-invariance nonlinearly realized} 
by the TD field $\chi=e^{\phi/F_\phi}$, with the spontaneous breaking 
of the scale invariance  characterized by the scale of $F_\phi$, 
while the Higgs (TD) mass term in the TD potential,  
on the  order of $m_F (\ll F_\phi)$,  
is the only source of the explicit breaking of the scale symmetry related (via PCDC) 
to the nonperturbative scale anomaly of the underlying theory.

One interesting  candidate for such technihadrons may be a resonance behind the diboson excess recently observed at the LHC 
at 2 TeV~\cite{Aad:2015owa, Khachatryan:2014hpa}, which can be identified with the walking technirho \cite{Fukano:2015hga}. 
The excesses suggest a characteristically small  width $\Gamma_{\rm total} 
<100$ GeV \cite{Aad:2015owa}, which can be naturally realized in the anti-Veneziano limit in Eq.(\ref{a-V:limit}): 
\beq
\frac{\Gamma_{\rm total}}{M_\rho}\simeq 
\frac{\Gamma(\rho \rightarrow WW/WZ)}{M_\rho} \simeq 
\frac{1}{48\pi}  \frac{g_{\rho\pi\pi}^2}{N_D} \sim \frac{1}{N_F N_C}\rightarrow 0\,,
\eeq 
where $N_D=N_F/2$ is the number of the weak-doublets. In fact our one-family model $N_F=8, N_C=4$ can
reproduces the features of the excesses very well \cite{Fukano:2015hga}. 
A smoking gun  of the walking technirho is the absence of the decay to the 125 GeV Higgs (TD), 
which is forbidden by the scale symmetry explicitly broken only by the Higgs (TD) mass term  
(corresponding to the nonperturbative scale anomaly in the underlying WTC)~\cite{Fukano:2015uga}.  
Actually, the salient feature of the scale symmetry of the generic effective theory 
containing the SM gauge bosons and the Higgs plus new vector bosons (any other massive particles as well), 
is the absence of the decay of the new vector bosons such as the technirho (and also other higher resonances) 
into  the 125 GeV Higgs plus the SM gauge bosons~\cite{Fukano:2015uga}. 
If such decays of new particles are not found at LHC Run II, 
then the 125 GeV Higgs is nothing but the dilaton 
(TD in the case of the WTC) 
responsible for the nonlinearly realized scale symmetry, i.e., 
the  spontaneous breaking of the scale symmetry, 
no matter what underlying theory may be beyond the SM. 
This should be tested in the  ongoing LHC Run-II.

\section{Conclusion}\label{conc}

The technidilaton, arising as a pseudo NG boson for spontaneous breaking of the scale symmetry 
in the walking techniicolor, can be identified with 
the Higgs discovered at the LHC. 
The smallness of the TD mass can be ensured by the characteristic feature of the walking, 
and the presence of the anti-Veneziano limit in the case of the 
large-flavor walking gauge theory including the one-family model. 
The couplings of the technidilaton to the standard model particles 
have so far been consistent with those of the LHC Higgs.  
Crucial deviation from the standard model Higgs will be found 
in the decay channel to bottom quark pair, which is significantly suppressed 
by the large decay constant compared to the electroweak scale, 
to be tested in the ongoing Run II or other future colliders.

Walking technipions, pseudo NG bosons for the chiral symmetry, 
get heavy due to the salient feature of the walking dynamics 
having the large anomalous dimension. The masses are lifted to be on the order of 
a few TeV scale, which is consistent with the current LHC-Run I limits, 
so the walking technipions may be uncovered soon, or more severely constrained 
at the ongoing Run II through each discovery channel.

Non-pseudo technihadrons in the walking technicolor cannot be so light, 
in contrast to the technidilaton. 
The masses are thus expected to be on the order of the scale of the scale symmetry breaking, 
TeV  scale, slightly higher than the electro weak scale. 
Among those Non-pseudos, the walking technirhos at around 2 TeV may be responsible for the 
recent excesses in the diboson channel, reported from the ATLAS group. 
The ongoing Run II will give the definite answer if it is the signal of the 
walking technirho.

\section*{Acknowledgments}

I would like to thank Hidenori S. Fukano, Masafumi Kurachi and 
Koichi Yamawaki for useful discussions and collaborations.  
This work was supported by 
the JSPS Grant-in-Aid for Scientific Research (S) \#22224003 and  
the JSPS Grant-in-Aid for Young Scientists (B) \#15K17645.

%\bibliographystyle{ws-procs975x65}
%\bibliography{ws-pro-sample}

\end{document}